\documentclass[fleqn,usenatbib]{mnras}
\usepackage{newtxtext,newtxmath}
\usepackage[T1]{fontenc}

\DeclareRobustCommand{\VAN}[3]{#2}
\let\VANthebibliography\thebibliography
\def\thebibliography{\DeclareRobustCommand{\VAN}[3]{##3}\VANthebibliography}

\usepackage{graphicx}	
\usepackage{amsmath}
\usepackage{xcolor}

\DeclareRobustCommand{\hlsout}{\bgroup\markoverwith{\textcolor{red}{\rule[.5ex]{2pt}{0.4pt}}}\ULon}
\DeclareRobustCommand{\hluline}{\bgroup\markoverwith{\textcolor{yellow}{\rule[-0.7ex]{2pt}{2pt}}}\ULon}

\title{A low mass and radius neutron star candidate in XTE J1810-189?}

\author[Shoutao Ban et al.]{Shoutao Ban,$^{1}$
Helei Liu,$^{1}$\thanks{E-mail: heleiliu@xju.edu.cn}
Zhaosheng Li,$^{2}$\thanks{E-mail:lizhaosheng@xtu.edu.cn}
Yupeng Chen,$^{3}$
Guoliang L\"{u},$^{1}$
Akira Dohi,$^{4,5}$
\newauthor{Tomoshi Takeda,$^{6}$
Hongbin Fan,$^{1}$
Chunhua Zhu$^{1}$
and Renxin Xu$^{7,8}$}\thanks{E-mail:r.x.xu@pku.edu.cn}
\\
$^{1}$School of Physical Science and Technology, Xinjiang University, Urumqi 830046, China\\
$^{2}$School of Science, Qingdao University of Technology, Qingdao 266525, P.R. China\\
$^{3}$Key Laboratory for Particle Astrophysics, Institute of High Energy Physics, Chinese Academy of Sciences, 19B Yuquan Road, Beijing 100049, China\\
$^{4}$Astrophysical Big-Bang Laboratory (ABBL), CPR, RIKEN, Wako, Saitama 351-0198, Japan\\
$^{5}$Interdisciplinary Theoretical and Mathematical Sciences Program (iTHEMS), RIKEN, Wako, Saitama 351-0198, Japan\\
$^{6}$Graduate School of Advanced Science and Engineering, Hiroshima University, 1-3-1 Kagamiyama, Higashi-Hiroshima, Hiroshima 739-8526, Japan\\
$^{7}$Department of Astronomy, Peking University, Beijing 100871, China\\
$^{8}$Kavli Institute for Astronomy and Astrophysics, Peking University, Beijing 100871, China}

\date{Accepted XXX. Received YYY; in original form ZZZ}
\pubyear{2025}

\begin{document}
\label{firstpage}
\pagerange{\pageref{firstpage}--\pageref{lastpage}}
\maketitle


\begin{abstract}

Photosphere radius expansion (PRE) bursts provide a crucial tool for constraining the mass and radius of neutron stars. In this study, we analyze time-resolved spectroscopic data from XTE J1810-189 in 2008, which exhibit evidence of a PRE event. We report here the possibility of a small-size and low-mass neutron star in XTE J1810-189 with use of the advantage of the direct cooling tail method. We obtained three sets of results, which can be broadly divided into high metal abundance (20 $\rm{Z}_{\odot}$ and 40 $\rm{Z}_{\odot}$), low metal abundance and hydrogen-rich (pure hydrogen, $\rm{Z}_{\odot}$, 0.3 $\rm{Z}_{\odot}$, 0.1 $\rm{Z}_{\odot}$, 0.01 $\rm{Z}_{\odot}$), and pure helium. In the high-metallicity scenario, the inferred neutron star mass is $<1.3\,M_{\odot}$ with a radius $<8\,\rm{km}$. In the low-metallicity, hydrogen-rich case, the mass ranges from 0.3 to 2.1 $M_{\odot}$ with radii of 7-13 km. For a pure-helium composition, we find two mass solutions: $1.08_{-0.22}^{+1.32}M_{\odot}$ (with $R>14\,\rm{km}$) and $2.5-2.9\,M_{\odot}$ (above the highest observed neutron star masses).
Additionally, we applied the touchdown method combined with an MCMC analysis, the results are consistent with those from the direct cooling tail method, but with a broader range. 
Our analysis of the time-resolved spectrum of burst suggests a high-metallicity atmosphere, but new observations are required to confirm this result.

\end{abstract}





\section{Introduction} \label{sec:intro}

Neutron stars (NSs), among the densest objects in the universe, possess central regions where densities exceed nuclear saturation density. The exact form of this highly dense matter remains uncertain \citep{2001AAS...19910501L,2019PhRvD.100l3010J}. 
Numerous equation of states (EOSs) have been developed based on various theoretical models~\citep{Baym_2018,2019PhRvC.100b5803S,2019PhRvL.122f1102B,2020PhRvD.101j3006E}. NSs derived from these EOSs can generally be categorized into two categories: gravity-bound NSs and self-bound NSs, depending on the nature of their surface binding forces ~\citep{1996cost.book.....G}. These two types exhibit contrasting behaviors at low masses: gravitationally bound NSs show an increase in radius with decreasing mass, while self-bound NSs exhibit the opposite trend (e.g., \citealp{1986ApJ...310..261A,1999PhRvC..60b5803G,doi:10.1142/S0218271810017597}). The measurement of NS mass and radius not only constrains the EOS for dense matter, but also offers valuable insights into the mechanisms of supernova explosions ~\citep{1994ApJ...425..195S} and the stellar evolution of low-mass X-ray binary stars (LMXBs)~\citep{2023Univ...10....3R}.

With the advancement of observational technologies, a variety of methods have been developed to measure the mass and radius of NSs through radio, X-ray, and gravitational wave observations~\citep{2024APh...15802935A}. For example, the mass of a NS can be accurately determined in double NS or NS-white dwarf (WD) systems using pulsar timing ~\citep{PhysRevLett.13.789, 2016ARA&A..54..401O,10.1093/mnras/stac3719}. The radius of a NS can be constrained by modeling the pulse profiles resulting from the non-uniform thermal surface emission of rotation-powered pulsars~\citep{1983ApJ...274..846P,2016ApJ...832...92O,2016RvMP...88b1001W}. In this approach, the periodic flux modulation of the NS is caused by the radiation flux change of hotspots on the distorted rotating NS surface under the influence of relativistic effects. This method is among the most promising avenues for obtaining robust constrains on NS parameters, and it has already yielded informative results for sources such as PSR J0030+0451 ~\citep{2019ApJ...887L..21R}, PSR J0740+6620 ~\citep{2021ApJ...918L..28M,2024ApJ...974..294S}, and XTE J1814-338~\citep{2024MNRAS.535.1507K}. The Neutron-star Interior Composition Explorer (NICER) is specifically designed to measure the NS radius from the observational thermal X-ray emission from the polar caps of X-ray pulsars~\citep{2012SPIE.8443E..13G,2014SPIE.9144E..20A}. Additionally, modeling the thermal emission from quiescent low-mass X-ray binaries(LMXBs)~\citep{2006ApJ...644.1090H}, estimating the gravitational redshift factor from the absorption edge energy produced in the NS photosphere ~\citep{2018ApJ...866...53L}, as well as conducting spectral studies from photospheric radius expansion (PRE) bursts all provide simultaneous measurements of NS mass and radius. 

 Over the past few decades, the X-ray telescopes (e.g., RXTE, NICER, HXMT...) have provided some of the best observations of type I X-ray burst to date~\citep{2020ApJS..249...32G,2022ApJS..260...25C,2024ApJ...975...67J,2024eas..conf.2072D}. A growing number of bursters, currently numbered at $\sim 120$\footnote{\url{https://sronpersonalpages.nl/~jeanz/bursterlist.html}}, offer a large enough sample to constrain the fundamental properties of NSs. Approximated 20\% of all bursts exhibit PRE, in which the energy release is sufficiently high for the flux at the NS surface to approach the local Eddington limit~\citep{1984ApJ...276L..41T,1984ApJ...277L..57L}. The Eddington flux and the apparent area of the emitting region derived from spectroscopic analysis of PRE bursts makes these events excellent candidates for measuring the masses and radii of the host NSs.

\cite{1986ApJ...305..246F} were the first to use the PRE burst data to determine the mass and radius of NS in MXB 1636-536. Later, \cite{2006Natur.441.1115O} determined the mass and radius of the NS in EXO 0748-676 using three key observable quantities: the Eddington flux $F_{\rm Edd}$, the redshift $z$, and the blackbody normalization $A$, which is related to the emitting area from the spectroscopic analysis of the PRE burst data. The large size and high mass of this NS strongly excluded the soft EOS of NS matter. Subsequently, \cite{2009ApJ...693.1775O} and \cite{2010ApJ...719.1807G} applied the same method to measure the masses and radii of NSs in globular clusters(whose distances were reasonably determined independently), employing the Bayesian method to account for the uncertainties in the observables. In addition to bursters in globular clusters,
\cite{2010ApJ...712..964G} measured the distance to 4U 1608-52 based on interstellar extinction, then determined the mass and radius of its NS. However, in these calculations, the color correction factor $f_{\rm c}$ was assumed to be constant, whereas it should vary as a function of the burst luminosity. Furthermore, the touchdown flux was assumed to be equal to the Eddington flux, which suffers uncertainties in determining whether the Eddington luminosity is reached at the moment of touchdown.

\cite{2011ApJ...742..122S} proposed an improved method using data from the cooling phase of the PRE burst. By fitting the data with an appropriate atmosphere model, their method allows for more reliable determination of the Eddington flux and blackbody normalization. This approach, however, depends on the chemical composition of the NS atmosphere model, which has been extensively studied over the past few decades~\citep{1987PASJ...39..287E,1991MNRAS.253..193P,2004ApJ...602..904M,2005A&A...430..643M}. \cite{2011A&A...527A.139S} computed a set of hot NS atmosphere models that account for variations in NS surface gravities, chemical compositions and relative luminosities. In consideration of metal absorption edges in the spectra, \cite{2015A&A...581A..83N} computed a detailed set of NS atmosphere model based on their previous models~\citep{2011A&A...527A.139S} with different exotic chemical composition that mimic the presence of burning ashes. \cite{2014MNRAS.442.3777P} found that the cooling tail method is only applicable when bursts are observed in the hard state at low mass accretion rates. Therefore, care must be taken in burst selection when using the cooling tail method. Subsequently, the mass and radius of NSs in low-mass X-ray binaries 4U 1720-429, 4U1724-307 and SAX J1810.8-260 were successfully constrained using the cooling tail method~\citep{2016A&A...591A..25N}. More recently, \cite{2017MNRAS.466..906S} proposed the direct cooling tail method and used it to measure the mass and radius of the NS in SAX J1810.8-2609. This improved method shifted the confidence regions by $1\sigma$ towards larger radii compared to the standard cooling tail method. \cite{2018ApJ...866...53L} and \cite{Fan_2026} applied the advantage of direct cooling tail method to determine the mass and radius of the NS in GRS 1747-312 and 2S 0918–549.

XTE J1810-189 is a transient source first discovered by the Rossi X-ray Timing Explorer ($RXTE$) in March 2008 ~\citep{2008ATel.1424....1M}. \cite{2008ATel.1443....1M} identified XTE J1810-189 as a NS through directional observations with $RXTE$ and placed an upper limit on its distance of 11.5 kpc, assuming its peak flux corresponds to the Eddington flux. \cite{2015MNRAS.450.2915W} conducted a detailed analysis of the type I X-ray burst data from XTE J1810-189, founding clear evidence of PRE and obtaining a distance range of 3.5--8.7 $\rm kpc$. In 2020, NICER observed further activity from XTE J1810-189. \cite{2023MNRAS.526.1154M} analyzed these NICER data and discovered that XTE J1810-189 features a moderately thick Comptonization region along with a broad iron K-shell emission line at $\sim$6.4 $\rm keV$. They calculated the blackbody radius during the outburst using distances provided by predecessors, and at a 68$\%$ confidence levels, the upper limit of the blackbody radius is less than 8 $\rm km$. They also noted that XTE J1810-189 exhibits several characteristics of ultracompact X-ray binaries (UCXBs), although definitive evidence is still lacking. However, the mass and radius of the NS in XTE J1810-189 remain undetermined in the literature. In this study, we focus on the time-resolved spectroscopy of PRE bursts observed from XTE J1810-189 to determine these critical parameters.

The structure of the paper is as follows.In section \ref{sec:Obs}, we analyze the spectrum of the burst. In section \ref{sec:cons}, we apply the direct cooling tail method to estimate the mass, radius and distance of the NS in XTE J1810-189. Additionally, we use MCMC simulations to combine observables and infer the mass and radius for comparison. In section \ref{sec:dis}, we discuss some of the issues this work are currently facing and the constrains on the EOS. Finally, the conclusions are summarized in section \ref{sec:con}.


\section{RXTE Observations and data analysis} \label{sec:Obs}

XTE J1810-189 underwent an outburst from March to September 2008, during which four type I X-ray bursts were detected by RXTE ~\citep{2008ATel.1443....1M,2015MNRAS.450.2915W}. Among them, the burst detected on 2008 May 4 (observation ID: 93433-01-06-02) exhibited PRE~\citep{2015MNRAS.450.2915W}. We adopted a hydrogen column density of $N_{\rm H} = 3.81^{+0.49}_{-0.46} 
\times 10^{22}\,\rm cm^{-2}$, consistent with the value reported by \cite{2015MNRAS.450.2915W}.

Since only PCU2 was active during the observation, all data were extracted from this unit. The data were analyzed using the \texttt{FTOOLS} software package version 6.13. The response matrix files were generated using \texttt{PCARSP}, and dead-time corrections for the spectra were applied following the recommendations of the RXTE team\footnote{\url{https://heasarc.gsfc.nasa.gov/docs/xte/recipes/pca_deadtime.html}}.
To accommodate the very low count rates, we used the Churazov weights ~\citep{1996ApJ...471..673C} for calculating $\chi^2$ in the fitting process, and a systematic error of 0.5\% was applied in the spectral analysis.

\subsection{Persistent Emission} \label{subsec:per}

Persistent emission spectra were extracted from the standard2 data using the standard criteria followed~\cite{2015MNRAS.450.2915W}. For the extraction, data from the top layer of PCU2 were selected, and X-ray emission associated with the type I X-ray burst was excluded (i.e. the 300 seconds before and 1000 seconds after the burst were ignored), resulting in an exposure time of 1880 seconds. A bright background model ($>40\,\rm count\,s^{-1}\,PCU^{-1}$) was used to create the background spectrum via the \texttt{pcabackest} program. The persistent spectrum in the range of 3-20$\,\rm keV$ was fitted with an absorbed power-law model, in addition to a gaussian component (\texttt{tbabs*(powerlaw+gau})). We attempted to add a blackbody model to detect blackbody component (\texttt{tbabs*(bbodyrad+powerlaw+gau})). 

\begin{figure}
\includegraphics[width=\columnwidth]{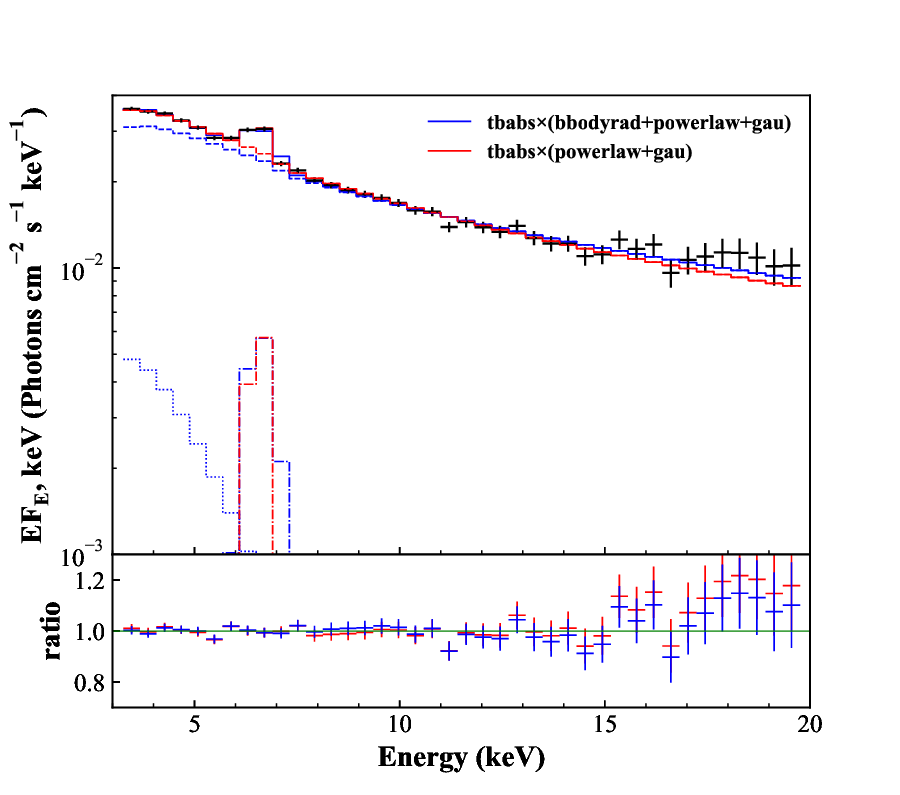}
\caption{Persistent emission spectrum of XTE J1810-189 (ObsID 93433-01-06-02). Red represents the \texttt{tbabs$\times$(powerlaw+gau)} model, and blue represents the \texttt{tbabs$\times$(bbodyrad+powerlaw+gau)} model. Solid lines, dash-dot line, dashed line, and dotted line correspond to the complete model, Gaussian, Powerlaw, and blackbody, respectively. The residuals related to the above two fitting models for each spectrum are also shown (bottom panel).}
\label{fig:per}
\end{figure}

Figure~\ref{fig:per} shows the fit to the persistent spectrum along with the residuals, and the relevant fitting parameters are given in Table  \ref{tab:efe}. For model without blackbody, the ratio of $\chi^2$ to degrees of freedom is 0.8, and this value becomes 0.76 after adding blackbody. The addition of the blackbody component reduces the power-law index by 0.1, which has little effect on the gaussian component. 
The contribution of the blackbody component to the energy spectrum is most evident at low energies, below 7 keV, and the main component of the spectrum is still power-law. The blackbody temperature is $0.77^{+0.18}_{-0.23}$ keV, and the normalization of the blackbody model is $14.50^{+44.51}_{-10.64}$. 
The large uncertainty in the blackbody normalization arises because the lower blackbody temperature reduces its contribution in the 3–7 keV band; this does not affect the conclusion that the power-law component dominates.
The fitting results of the model show that the persistent emission of XTE J1810-189 is consistent with the situation of 4U 1820-30 in the hard persistent states ~\citep{2017MNRAS.472.3905S}. We used cflux to calculate the flux in the 0.01-200 keV range. The flux of the persistent emission is $F$ = $(1.15 \pm 0.03)$ $\times10^{-9}\,\rm erg\,cm^{-2}\,s^{-1}$.

\begin{table}
\renewcommand{\arraystretch}{1.5}
\caption{Best-fitting parameters for persistent emission spectrum.}
\label{tab:efe}
\begin{tabular*}{\linewidth}{@{\extracolsep{\fill}} l p{2.2cm} c c @{}}
\hline
Model & Parameter & Model 1$^*$ & Model 2$^*$\\
\hline
\texttt{tbabs} & $N_{\rm H}$ & 3.81 & 3.81\\
\texttt{powerlaw} & Index & $2.04^{+0.02}_{-0.02}$ & $1.93^{+0.08}_{-0.09}$ \\
                  & norm  & $0.19^{+0.01}_{-0.01}$ & $0.15^{+0.03}_{-0.03}$ \\
\texttt{gau} & LineE (keV) & $6.53^{+0.13}_{-0.13}$ & $6.6^{+0.14}_{-0.13}$ \\
             & Sigma (keV) & $0.16^{+0.26}_{-0.15}$ & $0.35^{+0.25}_{-0.34}$ \\
             & norm $10^{-4}$ & $6.5^{+1.7}_{-1.5}$ & $9.0^{+3.6}_{-2.7}$ \\
\texttt{bbodyrad} & kT (keV) & - & $0.77^{+0.18}_{-0.23}$ \\
                  & norm     & - & $14.50^{+44.51}_{-10.64}$ \\
$\chi^2$/dof & {} & 31.33/39 & 25/33 \\
\hline
\multicolumn{4}{@{}p{\linewidth}@{}}{\footnotesize $^*$ Note: Models 1 and 2 represent \texttt{tbabs$\times$(powerlaw+gau)} and \texttt{tbabs$\times$(bbodyrad+powerlaw+gau)}, respectively.}
\end{tabular*}
\end{table}

\subsection{Fit of the burst spectra with an absorbed blackbody model} \label{subsec:per}

The time-resolved spectra were extracted from the science event file, while the background spectra were obtained from PCU2 in event mode using the same time settings as the persistent emission. Since the photon count rate for channel 11 is always 0, it was ignored in the fitting. During the fitting process, a significant deletion of the spectrum at channel 16 (around the 10 keV region) was observed. This phenomenon becomes more pronounced at higher photon count rates, but has minimal impact at low count rates. Given that the data were extracted from the PCA event mode data (E\_125us\_64M\_0\_1s) with energy channel allocation in M mode, the specifics of the data for this particular channel could not be determined. Since this detail is not critical to the work, channel 16 was also excluded from the fitting, which significantly improved the fit.

\begin{figure}
\includegraphics[width=\columnwidth]{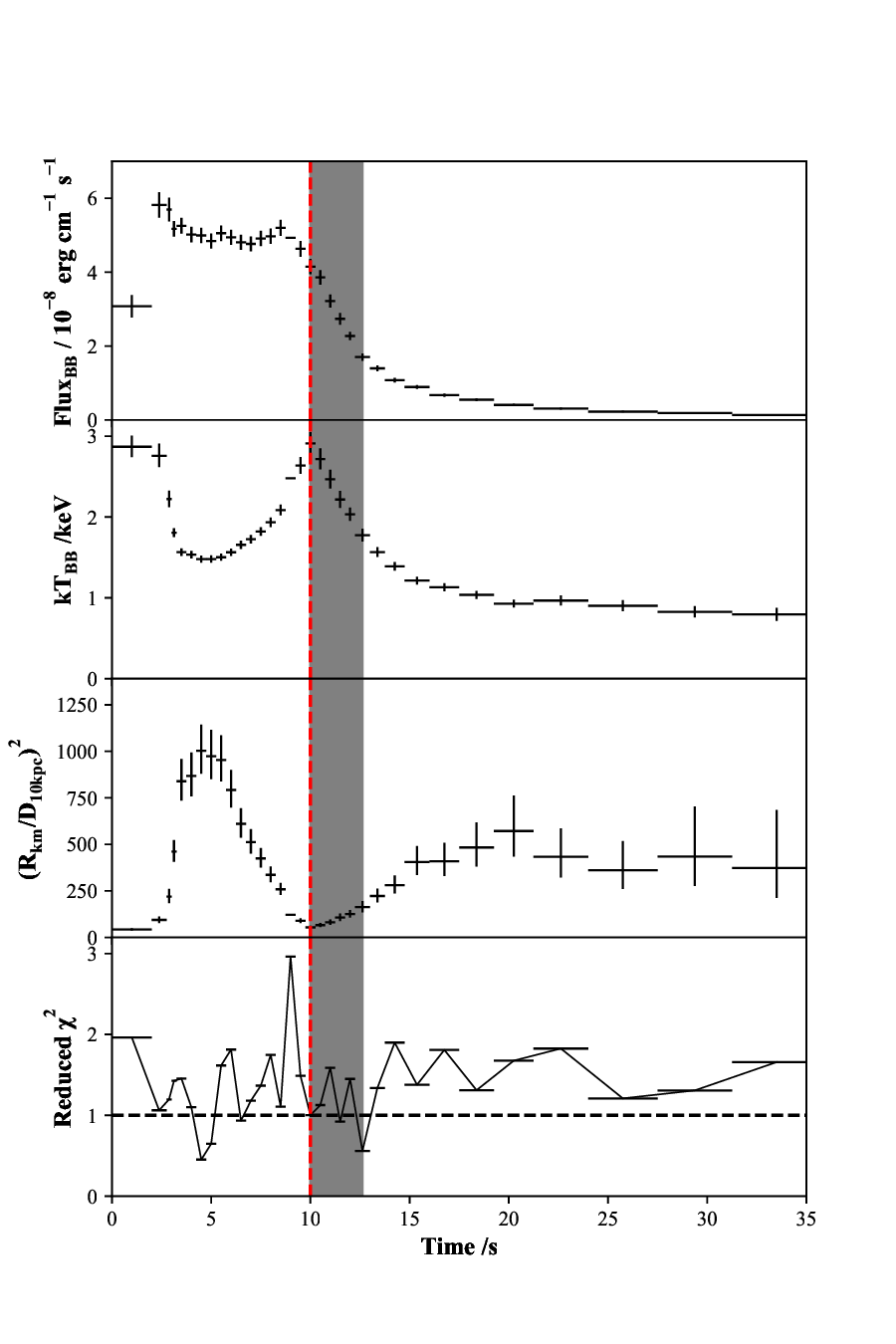}
\caption{Time-resolved spectra of PRE burst in XTE J1810-189 (ObsID 93433-01-06-02). The red dashed line labels the touchdown moment. The bold data points correspond to the fitting parameters at the touchdown moment.
\label{fig:PRE}}
\end{figure}

The burst spectra were fitted using an absorbed blackbody model (\texttt{tbabs*bbodyrad}) over the energy range of 3--20$\,\rm keV$. For the later stages of the burst, the fitting was concentrated in the range of 3--10$\,\rm keV$, as there were too few data point above 10 keV at this time. It is worth noting that a large $\chi^2/\rm dof$ value appears at $t$ = 9 s, likely due to poor data quality, as the spectrum at this time shows significantly anomalous dispersion compared to the other spectra.

The time-resolved spectra of the PRE burst were extracted with a time-bin size of 0.25\,s before the peak. To ensure statistical robustness of the data, the integration time was increased as the flux decreased, ensuring that the number of photons in each integration interval did not fall below 1000. The generated time-resolved spectra are presented in Figure~\ref{fig:PRE}, showing the thermal flux, blackbody temperature, blackbody normalization constant, and reduced $\chi^2$. The bolometric flux and its errors were estimated over the range of 0.01 - 200 keV using the \texttt{cflux}-model. The measured touchdown flux is $F_{\text{td}}$ = $(4.15 \pm 0.21)$ $\times10^{-8}\,\rm erg\,cm^{-2}\,s^{-1}$, with a corresponding peak flux of $F_P$ = $(5.82 \pm 0.21)$ $\times10^{-8}\,\rm erg\,cm^{-2}\,s^{-1}$, a blackbody temperature of $2.88^{+0.14}_{-0.13}$ $\rm keV$, and a normalized value of $53.98^{+9.10}_{-7.98}$ $( {\mathrm{km}}/{10~\mathrm{kpc}})^2$.


\section{The mass, radius and distance measurement of XTE J1810-189}\label{sec:cons}

\subsection{Direct cooling tail method}

The direct cooling tail method proposed by \cite{2017MNRAS.466..906S} provides more accurate confidence regions, fitting the observed dependence of the blackbody normalization on flux using an atmosphere model directly on the $M-R$ plane by interpolating theoretical dependencies for a given gravity $g$. This method allows us to obtain the best fitting parameters (mass, radius and the distance) by minimizing the function
\begin{equation}
\chi^2 = \sum_{i=1}^{N_{\mathrm{obs}}} \left[ \frac{(\omega \Omega - K_i)^2}{\Delta K_i^2} + \frac{\left(\omega f_c^4 \ell F_{\mathrm{Edd},-8} - F_{\mathrm{BB},i}\right)^2}{\Delta F_{\mathrm{BB},i}^2} \right]
\end{equation}
where $K_i$ and $F_{\mathrm{BB},i}$ are the observed blackbody normalization and flux during the cooling tail, and $\Delta K_i$ and $\Delta F_{\mathrm{BB},i}$ represent their errors. The parameters $\omega$, $f_c^4$, and $\ell$ are the theoretical spectral dilution factor, the color correction factor, and the relative luminosity ($L$/$L_{\mathrm{Edd}}$), respectively, as provided by the model. The Eddington flux $F_\mathrm{Edd}$ and the angular dilution factor $\Omega$ satisfy the following relationship:

\begin{equation}
F_{\mathrm{Edd}} = \frac{GMc}{k_{\mathrm{es}}D^2} \left(1 - \frac{2GM}{Rc^2}\right)^{1/2}
\end{equation}

and

\begin{equation}\label{eq:om}
\Omega = \left(\frac{R_{\rm km}\left(1 + z\right)}{D_{\mathrm{10kpc}}}\right)^{2}\,.
\end{equation}
where $\kappa_{\mathrm{es}} = 0.2(1+X) \, \mathrm{cm^2 \, g^{-1}}$ is the Thomson electron scattering opacity , with $X$ being the hydrogen mass fraction. 

\begin{figure}
\includegraphics[width=\columnwidth]{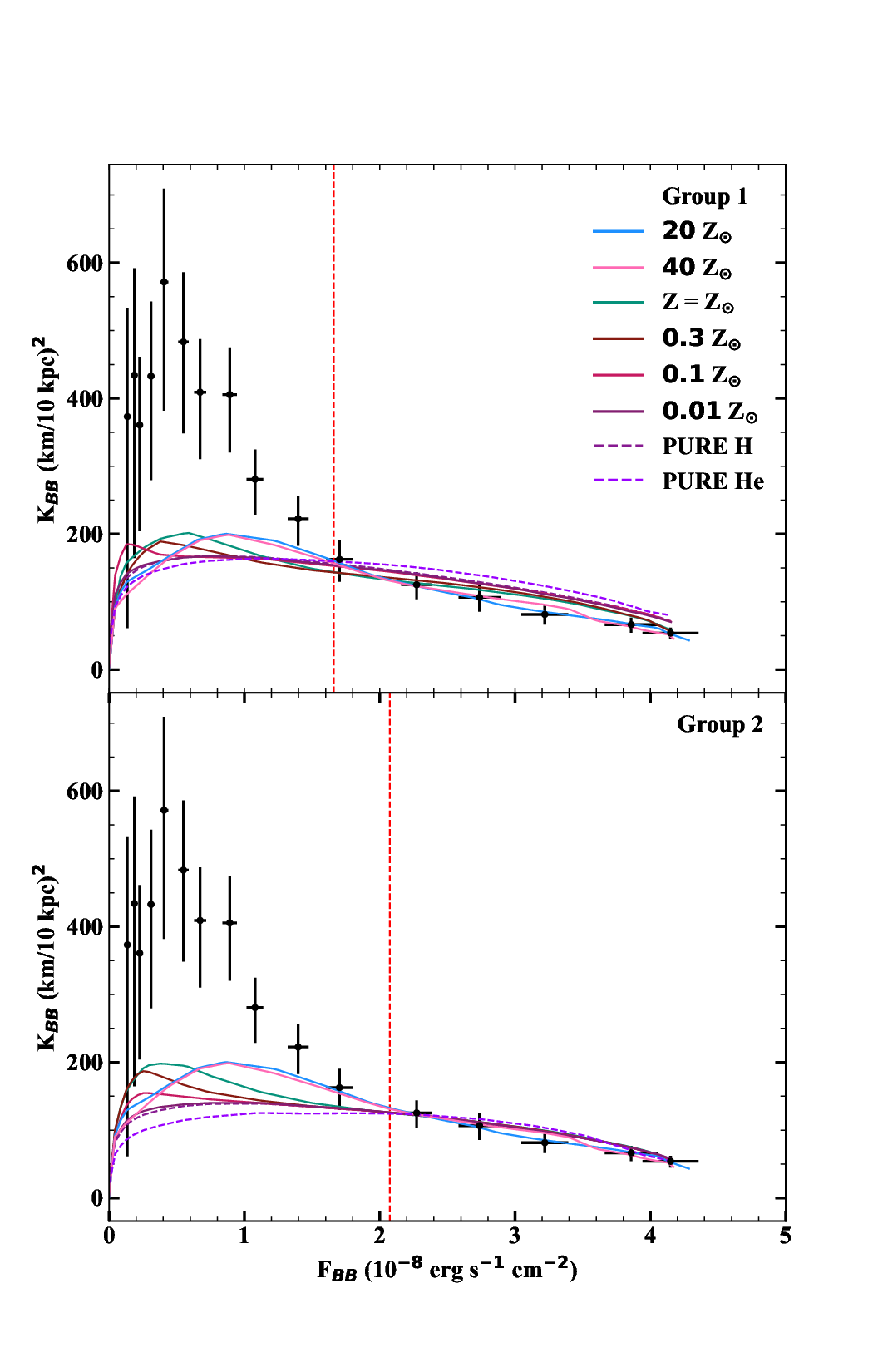}
\caption{Fitting between the cooling tail data and the theoretical curve of the PRE burst. The black dots marked with error bars represent the observed {$K$} - $F_{\text{BB}}$. The upper figure shows the results of Group 1, where the red dashed line represents $F$ = 0.4 $F_{\text{td}}$, and the data points on the left are not considered during the fitting process. The lower figure represents the results of Group 2, where the red dashed line represents $F$ = 0.5 $F_{\text{td}}$.} The curves represent the theoretical model $w\Omega$ - $wf_c^4\ell$$F_{\mathrm{Edd},-8}$, with pink and blue corresponding to the atmospheric model of $40$ $\rm{Z}_{\odot}$ and $20$ $\rm{Z}_{\odot}$~\citep{2015A&A...581A..83N}, respectively. Green solid line, brown solid line, magenta solid line, purple solid line, purple dashed line, and bright purple dashed line respectively correspond to $\rm{Z}_{\odot}$, 0.3 $\rm{Z}_{\odot}$, 0.1 $\rm{Z}_{\odot}$, 0.01 $\rm{Z}_{\odot}$, pure H, and pure He models~\citep{2011A&A...527A.139S}.
\label{fig:fit}
\end{figure}

\begin{table*}
\renewcommand{\arraystretch}{1.5}
\caption{The best fitting parameters ($D$, $M$, $R$, $\Omega$ and $F_{\rm Edd}$) and the range of 99$\%$ confidence levels for $M$, $R$, and $D$. $X$, $Y$ and $Z$ are the mass fractions of hydrogen, helium, and metal, respectively.}
\label{tab:fittable}
\begin{tabular}{lcccccccccc}
\hline
Group & Model & $X$ & $Y$ & $Z$ & $D$ & $M$ & $R$ & $\Omega$ & $F_\mathrm{Edd}$ & $\chi^2$\\
{} & {} & {} & {} & {} & $\rm kpc$ & $\rm{M}_{\odot}$ & $\rm km$ & $(\rm km/10\,kpc)^{2}$ & $\times10^{-8}\,\rm erg\,cm^{-2}\,s^{-1}$ & {}\\
\hline
{} & $\rm{Z}_{\odot}$ & 0.7374 & 0.2492 & 0.00134 & 2.3$^{a}$ & 0.18$^{a}$ & 5.59$^{a}$ & 652 & 3.89 & 50.46\\
{} & $0.3$ $\rm{Z}_{\odot}$ & 0.550 & 0.185 & 0.256 & 3.8$^{a}$ & 0.51$^{a}$ & 9.13$^{a}$ & 691 & 3.88 & 86.08\\
{} & $0.1$ $\rm{Z}_{\odot}$ & 0.550 & 0.185 & 0.256 & 6.0$^{a}$ & 2.05$^{a}$ & 8.9$^{a}$ & 688.5 & 3.87 & 105.87\\
{1} & $0.01$ $\rm{Z}_{\odot}$ & 0.550 & 0.185 & 0.256 & 6.1$^{a}$ & 2.10$^{a}$ & 9.08$^{a}$ & 699.42 & 3.82 & 97.54\\
{} & $20$ $\rm{Z}_{\odot}$ & 0.550 & 0.185 & 0.256 & $3.0^{+2.5}_{-0}$ & $0.30^{+0.79}_{-0.02}$ & $5.10^{+3.14}_{-0.33}$ & 335 & 4.00 & 1.80\\
{} & $40$ $\rm{Z}_{\odot}$ & 0.352 & 0.118 & 0.530 & $4.6^{+1.55}_{-1.11}$ & $0.65^{+0.65}_{-0.30}$ & $6.14^{+1.86}_{-1.14}$ & 262 & 3.97 & 3.34 \\
{} & pure H & 1 & 0 & 0 & 6.0$^{a}$ & 2.10$^{a}$ & 10.36$^{a}$ & 743.11 & 3.86 & 103.05\\
{} & pure He & 0 & 1 & 0 & 9.3$^{a}$ & 2.99$^{a}$ & 12.26$^{a}$ & 621.66 & 3.82 & 184.65\\
\hline
{} & $\rm{Z}_{\odot}$ & 0.7374 & 0.2492 & 0.00134 & $4.3^{+2.2}_{-0.9}$ & $0.67^{+1.33}_{-0.26}$ & $9.51^{+2.89}_{-2.00}$ & 618 & 3.88 & 15.43\\
{} & $0.3$ $\rm{Z}_{\odot}$ & 0.7374 & 0.2586 & 0.004 & $6.4^{+0.3}_{-3.4}$ & $1.73^{+0.36}_{-1.31}$ & $12.09^{+0.76}_{-5.09}$ & 618 & 3.86 & 14.15\\
{} & $0.1$ $\rm{Z}_{\odot}$ & 0.7374 & 0.2613 & 0.256 & $6.4^{+0.3}_{-3.4}$ & $1.73^{+0.36}_{-1.41}$ & $12.16^{+0.76}_{-5.16}$ & 623 & 3.87 & 13.6\\
2 & $0.01$ $\rm{Z}_{\odot}$ & 0.7374 & 0.2625 & 0.000134 & $6.5^{+0.2}_{-3.6}$ & $2.01^{+0.11}_{-1.72}$ & $10.84^{+2.17}_{-3.84}$ & 615 & 3.85 & 13.17\\
{} & $20$ $\rm{Z}_{\odot}$ & 0.550 & 0.185 & 0.256 & $2.1^{+3.0}_{-1.3}$ & $0.14^{+0.83}_{-0.12}$ & $3.60^{+4.15}_{-2.25}$ & 332 & 4.03 & 1.07\\
{} & $40$ $\rm{Z}_{\odot}$ & 0.352 & 0.118 & 0.530 & $4.2^{+1.6}_{-3.6}$ & $0.54^{+0.65}_{-0.53}$ & $5.63^{+1.92}_{-4.79}$ & 261 & 4.01 & 4.57 \\
{} & pure H & 1 & 0 & 0 & $5.6^{+0.3}_{-2.4}$ & $1.49^{+0.41}_{-1.09}$ & $11.13^{+0.56}_{-3.88}$ & 653 & 3.86 & 13.55\\
{} & pure He$^{b}$ & 0 & 1 & 0 & $7.3^{+2.7}_{-0.8}$ & $1.08^{+1.32}_{-0.22}$ & $15.45^{+2.99}_{-1.34}$ & 564 & 3.77 & 16.73\\
{} & pure He$^{b}$ & 0 & 1 & 0 & $8.7 - 10.1$ & $2.53 - 2.89$ & $10.59 - 15.42$ & - & - & -\\
\hline
\multicolumn{11}{l}{\footnotesize $^a$ - The confidence interval is not provided because $\chi^2$ is too large and the range is too small.}\\
\multicolumn{11}{l}{\footnotesize$^b$ - The results of the pure He model have two parts, with the same best fitting parameters.}\\
\end{tabular}
\end{table*}

According to the report by \cite{2010arXiv1004.4086C}, the accreted material of XTE J1810-189 has a low abundance of hydrogen. We used the atmosphere model of \cite{2015A&A...581A..83N} and \cite{2011A&A...527A.139S}, which provides the variation of $f_{\rm c}$ with $\ell$ for three surface gravities and four metal abundances, supporting hydrogen-poor conditions in the metal abundance settings. We investigated a grid of models with masses ranging from $0.1$ to $3\,\rm{M}_{\odot}$ in steps of $0.01\,\rm{M}_{\odot}$ and radii spanning from $0.1$ to $18\,\text{km}$ in steps of $0.01\,\text{km}$. For each pair of mass-radius, we interpolated the atmospheric model by calculating its surface gravity $\log g$ to obtain the theoretical curve $\omega$ - $\omega f_c^4$. Following the approach of \cite{2017MNRAS.466..906S}, when $\ell$ is less than 0.8, we used the relationship $\omega f_c^4$ - $\ell$ and $\omega$ - $\ell$ for interpolation (treating $\omega$ and $f_c$ as functions of $\ell$). We simply extend the $g_\mathrm{rad}/g$ grid and switch to using the relationship $f_c$ - $g_\mathrm{rad}/g$ for interpolation when $\ell$ is greater than 0.8.

We do not directly calculate the minimized $\chi^2$ and its corresponding values of $D$ for each pair of $M$ and $R$ as in \cite{2017MNRAS.466..906S}. Instead, we calculate a grid of $D$ values ranging from 0.1 to 9 kpc with a step size of 0.1 kpc, based on the range of $D$ suggested by \cite{2015MNRAS.450.2915W}. This approach allows us to determine the range of $D$ for each pair of $M$ and $R$. In our calculations, parameter sets that do not satisfy the causal condition $R\,>\,2.9GM/c^2$ are excluded ~\citep{2015arXiv151006962S}. The cooling tail data used for fitting are represented by the gray area in Figure~\ref{fig:PRE}, with a minimum flux limit set to 0.4 $F_{\mathrm{Edd}}$. Due to uncertainties in the low-luminosity region, a lower limit is not considered, though this could provide additional data points. Considering the sensitivity of the direct cooling tailing method to data selection, we also calculated a case with a lower limit of $0.5\,F_{\mathrm{Edd}}$. For convenience, we refer to the case with a lower limit of $0.4\,F_{\mathrm{Edd}}$ as Group 1, and the case with a lower limit of 0.5 $F_{\mathrm{Edd}}$ as Group 2. We did not consider the pure iron model for both groups, as it is not realistic. 

\begin{figure*}
\includegraphics[width=\textwidth]{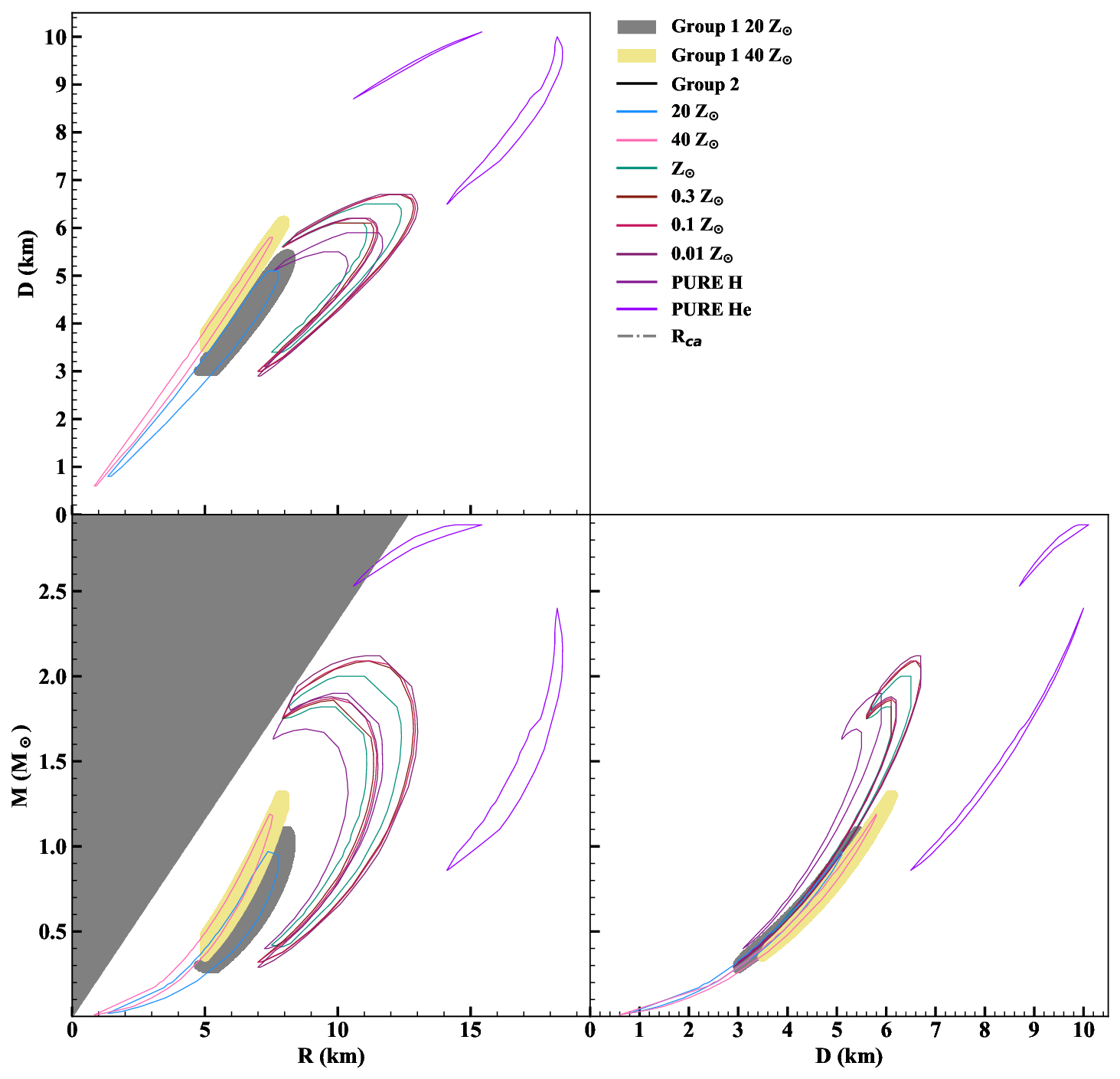}
\caption{Constraints on $M$, $R$, and $D$ obtained from direct cooling. The color mapping is consistent with Figure \ref{fig:fit} for models with different atmospheric composition. The curves represent the 99$\%$ confidence levels boundaries, with color blocks indicating the Group 1 and solid lines indicating to the Group 2. The gray dashed line represents the causality condition ($R_c = 1.45 R_S$), with the non-physical region above it shaded in gray.}
\label{fig:all_confidence}
\end{figure*}

The best-fitting parameters are presented in Table \ref{tab:fittable}, and the fitting relationship between the data and the theoretical model is shown in Figure \ref{fig:fit}. For all models with acceptable $\chi^2$ values (below 20), the confidence regions of M, R, and D are determined by $\Delta \chi^2$. The parameter ranges with 99$\%$ confidence for M, R, and D are also provided in Table \ref{tab:fittable}, and the distribution of parameters is plotted in Figure \ref{fig:all_confidence}. 

For Group 1, the minimum $\chi^2$ values (for 4 dof) for the 20 $\rm{Z}_{\odot}$ and 40 $\rm{Z}_{\odot}$ atmospheric models are 1.80 and 3.34, respectively. Their mass and radius distributions are very similar, with radii generally ranging from 4.8 to 8.2 km and masses from 0.28 to 1.3 $\rm{M}_{\odot}$. Models with pure H, pure He and solar metal abundance exhibit excessively large $\chi^2$ values, suggesting that low metal abundance is not preferred.

For Group 2,  the degrees of freedom are 3. The minimum $\chi^2$ values for the 20 $\rm{Z}_{\odot}$ and 40 $\rm{Z}_{\odot}$ atmospheric models are 1.07 and 4.57, respectively. The results of the high metal abundance in group 2 are similar to those in group 1, but with a larger range, mass less than 1.2 $\rm{M}_{\odot}$, radius less than 7.8 km. The $\chi^2$ values of all low metal abundance models are between 10 and 20. The mass and radius of the remaining pure H, $\rm{Z}_{\odot}$, 0.3 $\rm{Z}_{\odot}$, 0.1 $\rm{Z}_{\odot}$, and 0.01 $\rm{Z}_{\odot}$ models are concentrated between 7 to 13 km and 0.3 to 2.1 $\rm{M}_{\odot}$. The results of the pure He model are divided into two parts, where the mass and radius of the region with the higher mass at the top are $2.89^{+0.85}_{-0.19}$ $\rm{M}_{\odot}$ and $15.45^{+2.78}_{-1.27}$ km, and the mass and radius of the region with less mass below are $1.08^{+1.32}_{-0.22}$ $\rm{M}_{\odot}$ and $15.45^{+2.92}_{-1.34}$ km.

\subsection{
MCMC simulations of combining observables to infer mass and radius of NS} \label{subsec:MR}

In a PRE burst, the touchdown flux $F_{\rm td}=\frac{1}{4\pi D^2}\frac{4\pi GMc}{\kappa_{\rm es}}(1-\frac{2GM}{Rc^2})^{1/2}$ and the blackbody normalization $A=f_{\rm c}^{-4}\frac{R^2}{D^2}(1-\frac{2GM}{Rc^2})^{-1}$ can be combined to estimate the mass and radius of the NS (this method is known as the ``touchdown method"). Here, the distance to the target $D$ is crucial for determining $M$ and $R$. We use the same distance range as the direct cooling tail ($\sim 0.1-9\,\rm kpc$).

The Bayesian approach is adopted to constrain the mass and radius of the NS in XTE J1810-189 following~\cite{2016ApJ...820...28O}. Based on Bayes' theorem, the likelihood

\begin{equation}
 P(M,R|{\rm data})=CP({\rm data}|M,R)P_{\rm prior}(M)P_{\rm prior}(R)
\end{equation}
where $P_{\rm prior}(M)$ and $P_{\rm prior}(R)$ are the priors over the mass ($M$) and radius ($R$), respectively, and $C$ is a constant. Assuming that $F_{\rm td}$ and $A$ are ideally uncorrelated measurements, we can write
\begin{equation}
\begin{split}
 P({\rm data|M,R})=& \int P(D)dD \int p(f_{\rm c})df_{\rm c} \int P(X)dX\\
 &\times P[F_{\rm td}(M,R,D,X)]P[A(M,R,D,f_{\rm c})]
\end{split}   
\end{equation}
where $P(F_{\rm td})$ and $P({A})$ are the posterior likelihoods of the observed touchdown flux and the blackbody normalization, respectively. $P(D)$, $P(X)$, and $P(f_{\rm c})$ represent the priors over the distance, hydrogen mass fraction, and color correction factor of the NS, respectively. We employ MCMC methods\footnote{The open-source PYTHON package EMCEE (V2.2.1) was used: \url{https://emcee.readthedocs.io/en/v2.2.1/}} to sample the probability distributions across the parameter space.

\begin{figure}
\includegraphics[width=\columnwidth]{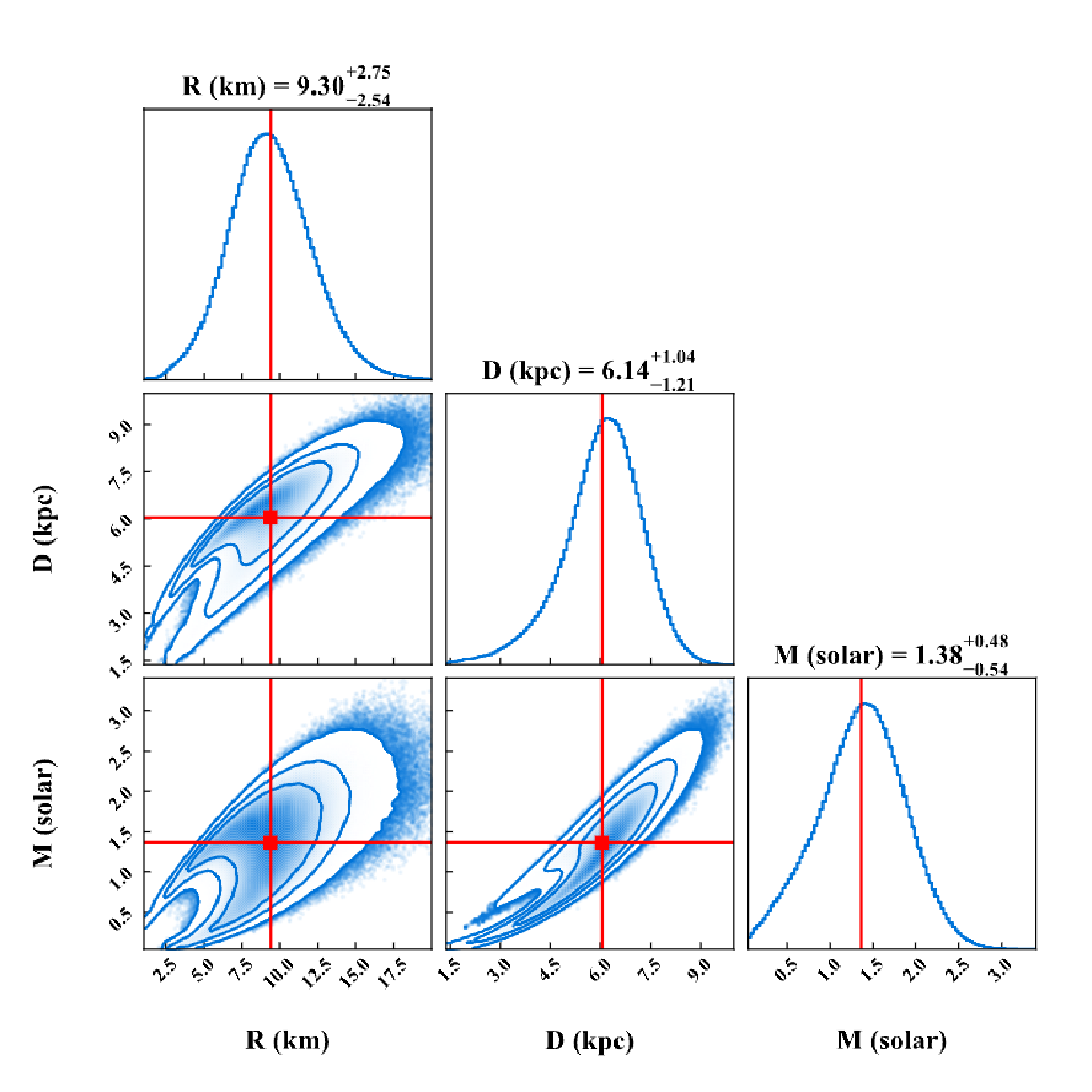}
\caption{Posterior distribution of Group 1 for the mass, radius and distance parameters from the MCMC simulations, the contour lines from inside to outside represent confidence intervals of 68\%, 90\%, and 99\%. The top tag displays the median, with its error corresponding to the 68\% confidence level. The outermost points correspond to the marginal distribution of the parameters. Along the diagonal are the marginalized posteriors. The maximum likelihood estimates for the posteriors are indicated at the intersection of the two red lines.}
\label{fig:mcmc}
\end{figure}

We selected flat priors for the NS mass, radius, distance, with the following ranges: $M\sim0.1-3.5\,\rm{M}_{\odot}$, $R\sim0.1-20\,\rm km$, respectively. The values of $F_{\text{td}}$ = ($4.15 \pm 0.21$) $\times10^{-8}\,\rm erg\,cm^{-2}\,s^{-1}$ and $A$ = $53.98^{+9.10}_{-7.98}$ $( {\mathrm{km}}/{10~\mathrm{kpc}})^2$, derived in section \ref{subsec:per}, were used for the calculation. We assume Gaussian distributions for both quantities and adopt the square of the error as the standard deviation. The error of $A$ was simply taken as the average error of 8.54. The color correction factor $f_c$ increases significantly at the touchdown moment, rising from approximately 1.75 to 1.9 as the atmospheric composition changes from pure He to pure H~\citep{2011A&A...527A.139S,2012A&A...545A.120S}. Secondly, according to \cite{2015A&A...581A..83N}, for high metal abundance, $f_c$ varies from 1.55 to 1.73 with atmospheric composition from 20 $\rm{Z}_{\odot}$ to 40 $\rm{Z}_{\odot}$. Considering the effect of changes in $f_c$ and the fact that an overly broad X range can lead to overdivergent results, we consider three combinations of $f_c$ and X: (i)for high metal abundance, $X=0.35-0.55$ and $f_{\rm c}=1.55-1.73$; (ii) for low metal abundance, $X=0.35-0.55$ and $f_{\rm c}=1.75-1.90$; and (iii) for an H-rich atmosphere, $X=0.7-1.00$ and $f_{\rm c}=1.75-1.90$.

\begin{table}
\renewcommand{\arraystretch}{1.5}
\caption{The posterior distributions of M, R, and D with three sets of parameter settings, at the 90\% confidence level.}
\label{tab:mcmc}
\begin{tabular*}{\linewidth}{@{\extracolsep{\fill}} l c c c c c @{}} 
\hline
Group & X & $f_c$ & M & R & D\\
{} & {} & {} & $\rm{M}_{\odot}$ & $\rm km$ & $\rm kpc$ \\  
\hline 
1 & 0.35-0.55 & 1.55-1.73 & $1.38^{+0.68}_{-0.90}$ & $9.30^{+5.36}_{-6.29}$  & $6.14^{+5.36}_{-6.29}$\\
2 & 0.35-0.55 & 1.75-1.90 & $2.02^{+0.95}_{-1.31}$ & $14.08^{+4.79}_{-5.96}$ & $7.45^{+1.59}_{-2.65}$\\ 
3 & 0.7-1.0   & 1.75-1.90 & $1.66^{+0.76}_{-1.04}$ & $11.21^{+4.36}_{-4.76}$ & $5.97^{+1.33}_{-2.08}$\\
\hline
\end{tabular*}
\end{table}

\begin{figure}
\includegraphics[width=\columnwidth]{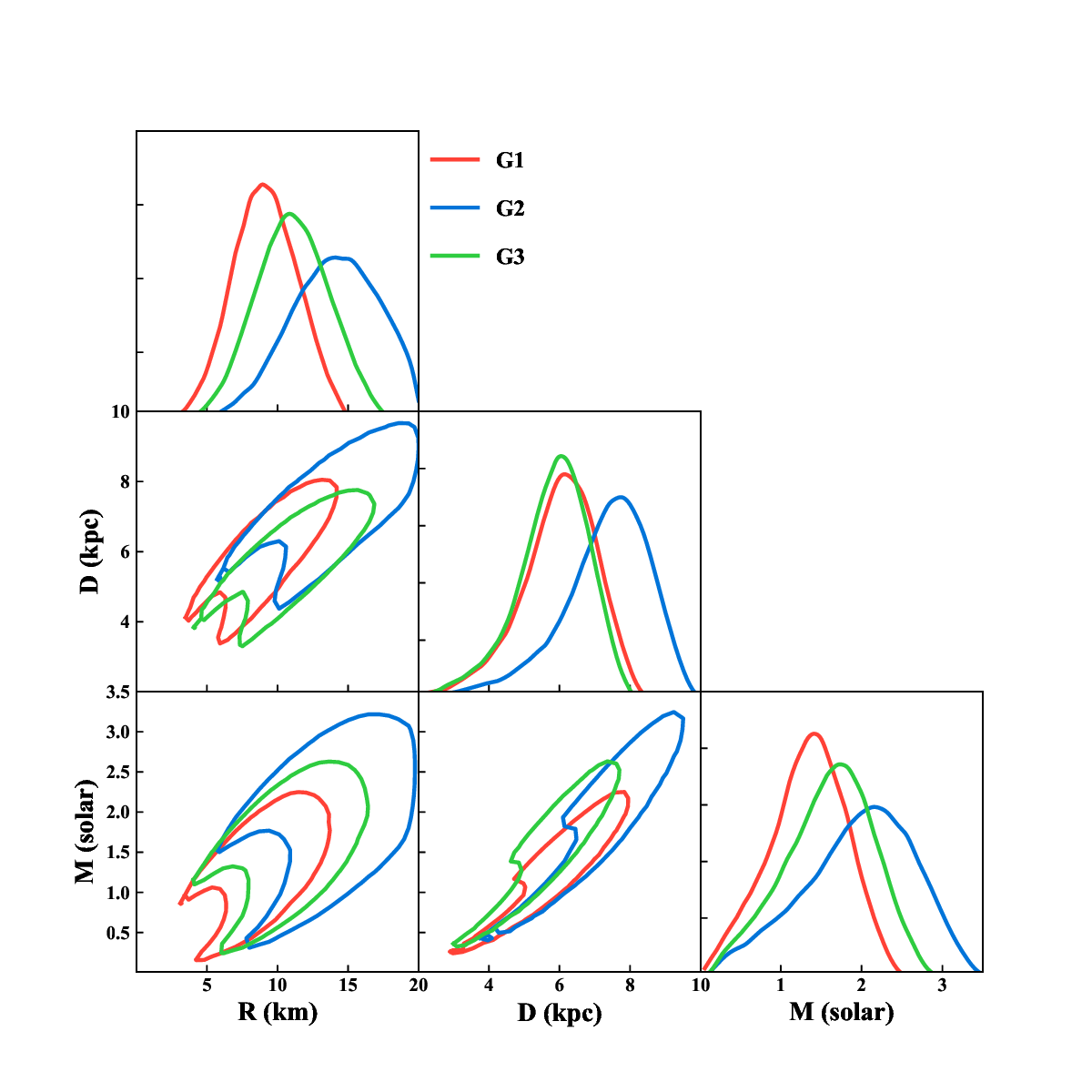}
\caption{The posterior distributions of M, R, and D with three sets of parameter settings, at the 90\% confidence level.}
\label{fig:all_mcmc}
\end{figure}

The posterior distributions of Group 1 for the mass ($M$), radius ($R$), and distance ($D$) for the first group are shown in Figure \ref{fig:mcmc}, where only one is plotted as an example, and the distribution of other combinations are listed in Table \ref{tab:mcmc}. In Figure \ref{fig:mcmc}, the contour lines correspond to 68\%, 90\%, and 99\% confidence regions, and the marginal posteriors for each parameter are displayed along the diagonal. The maximum likelihood estimate is indicated by the intersection of the two red lines. In Figure \ref{fig:all_mcmc}, the 90\% confidence intervals for the three groups are plotted, which clearly show: (i) an increase in $f_c$ leads to an increase in the mass and radius of NS; (ii) an increase in X reduces the NS's mass and radius, which is consistent with the case of cooling tails. The results indicate that under rich He conditions, XTE J1810-189 may have a neutron star with intermediate or large mass; in other cases, it corresponds to a neutron star with intermediate or low mass.


\section{DISCUSSION} \label{sec:dis}

Up to now, observations of XTE J1810-189 remain limited, with only a single PRE burst available for constraining its mass and radius. Here, we discussed the possible reasons, as well as the theoretical limitations on the current results. 

The results of the direct cooling tail show three types: (a) neutron stars with extremely small mass and radius, whose atmosphere has high metal abundance; (b) conventional neutron stars, whose atmosphere has low metal abundance and hydrogen-rich; (c) Neutron stars with large masses and radii with a pure helium atmosphere. For convenience, (a), (b) and (c) are simply used to represent the three results obtained using the direct cooling tail, and (d) to represent the results obtained by the ``touchdown method".

\subsection{Atmospheric composition}  \label{subsec:atm}

In the direct cooling tail method, both Group 1 and Group 2 tend to atmospheres with high metal abundance, and give extremely small mass and radius. However, \cite{2015A&A...581A..83N} points out that for atmospheric models with high metal abundance, the absorption edge of Fe ions strongly affect the emerging radiation at low relative luminosity. \cite{2017MNRAS.464L...6K} analyzed a PRE outburst of HETE J1900.1–2455 and found that the time-resolved spectra showed a sudden increase in $\chi^2$ in two regions: PRE phase and after the cooling tail. They improved the fit by adding the \texttt{edge} model and obtained the edge energies at these two regions as $8.07 \pm {0.08}$ keV and $7.62 \pm {0.09}$ keV, which is consistent with the H-like Fe ions after accounting for redshift. \cite{2018ApJ...866...53L} analyzed similar phenomena in GRS 1747–312, where the edge energies changed from 9 keV to 8 keV during the PRE phase and cooling tail phase. They showed that both energy levels correspond to the hydrogen-like Ni edge, and the change in edge energy is caused by the contraction of the photosphere. At the photosphere surface far from the NS, the redshift effect is minor, but as the photosphere surface approaches the NS surface, the redshift effect increases, leading to the observed decrease in edge energy.

\begin{figure}
\includegraphics[width=\columnwidth]{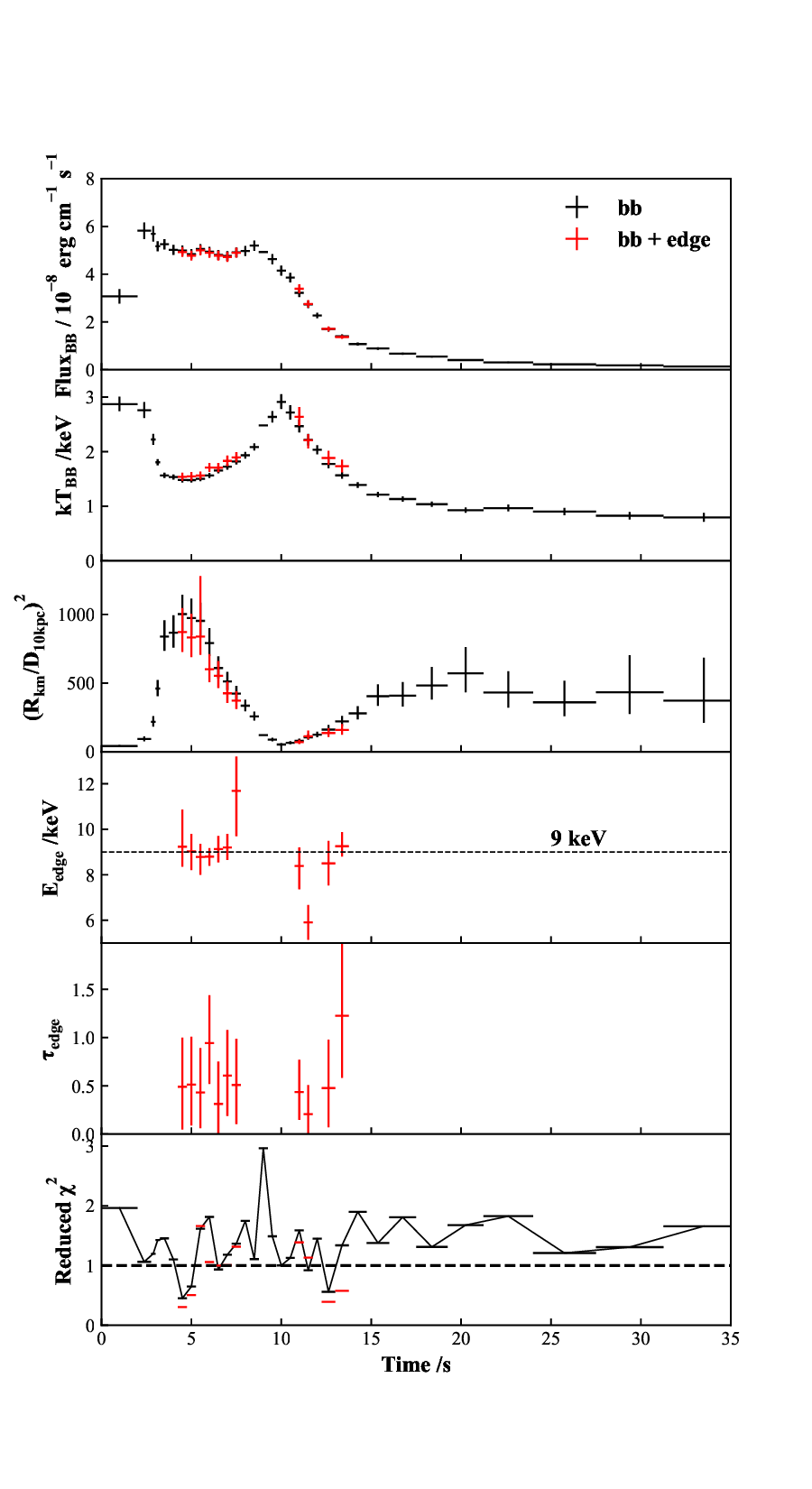}
\caption{The black data points are the same as in Figure~\ref{fig:PRE}; the results of the \texttt{bb} + \texttt{edge} model plotted in red, and the edge energy at 9 keV is marked with a black dashed line.}
\label{fig:edge}
\end{figure}

To test the validity of (a), we added the \texttt{edge} model to refit the entire time-evolution spectrum, And plot the new results together with the old ones in Figure~\ref{fig:edge}. Similar to the cases of \cite{2017MNRAS.464L...6K} and \cite{2018ApJ...866...53L}, the edge model can only provide reasonable results during the PRE phase and cooling tail phase. Edge energy is close to 9 keV in the PRE phase, and it only slightly decreases from the PRE phase to the cooling tail phase.We adopt the same method as \cite{2018ApJ...866...53L}, assuming the observed edge energy comes from the hydrogen-like Ni at 10.8 keV edge, and estimate the redshift based on this. For the PRE period and touchdown period, we selected the first 6 points and two points near 8 keV, respectively, for calculating the average edge energy $E_{\rm{edge,pre}}$ = $9.03^{+0.67}_{-0.75}$ keV and $E_{\rm{edge,td}}$ = $8.44^{+0.99}_{-0.90}$ keV. The ratio of the energy can be used to calculate the redshift (1+$z_{\rm {pre}}$) = $1.17^{+0.11}_{-0.08}$, (1+$z_{\rm {td}}$) = $1.26^{+0.15}_{-0.13}$ and the ratio of the two is (1+$z_{\rm{td}}$)/(1+$z_{\rm {pre}}$) = $1.08^{+0.21}_{-0.19}$. We calculate the average flux $F_{\rm{pre}}$ = $4.99 \pm {0.20}$ $\times10^{-8}\,\rm erg\,cm^{-2}\,s^{-1}$ in the PRE phase to obtain $F_{\rm{pre}}$/$F_{\rm{td}}$ = $1.20^{+0.12}_{-0.10}$, which satisfy the relation $F_{\rm{pre}}$(1+$z_{\rm{pre}}$) = $F_{\rm{td}}$(1+$z_{\rm{td}}$) within the error range. The above results indicate that (a) is acceptable, but we must point out that the addition of the \texttt{edge} model did not improve the fit, and the detected edge energy may even be due to poor data quality and low energy resolution. To validate this possibility, new high-quality PRE observations of XTE J1810-189 are required.

\subsection{The particularity of the source}  \label{subsec:particularity}

Regarding the scope of application for the direct cooling tail method, \cite{10.1093/mnras/stu2073} has conducted a detailed discussion and provided some criteria for judgement. They require that the source burst be in a hard state, with the flux of persistent emission less than about 3$\%$ of the Eddington flux, and the $K_{\rm{td}/2}/K_{\rm{td}}$ between 2 and 3.5, where $K_{\rm{td}}$ and $K_{\rm{td}/2}$ represent the blackbody normalization at the moment of touchdown and the blackbody normalization when the flux is reduced to half of the $F_{td}$, respectively. \cite{2015MNRAS.450.2915W} has determined that the burst is in the hard state by the highly variable performance of the burst in the colour-colour diagram (CCD) (see Fig.~2 in \cite{2015MNRAS.450.2915W} for detail). For XTE J1810-189, $K_{\rm{td}/2}/K_{\rm{td}}$ is 3.01 when $F_{\rm{td}/2}/F_{\rm{td}}$ equals 0.41. Using the Eddington flux obtained from the direct cooling tail, the proportions of the flux of persistent emission for the $20$ $\rm{Z}_{\odot}$ and $40$ $\rm{Z}_{\odot}$ models are 0.295 and 0.293, respectively. All of these conditions indicate that the PRE burst of XTE J1810-189 in 2008 meets the conditions for prediction using the neutron star atmosphere model.

We note that \citet{10.1093/mnras/stu2073} reported that UCXBs exhibit X-ray behavior distinct from other hard-state sources. 
If XTE J1810–189 is indeed an UCXB, as proposed by \cite{2023MNRAS.526.1154M}, this could significantly affect our conclusions.
In particular, \citet{2023MNRAS.526.1154M} argued that the burst spectra and durations in XTE J1810-189 are consistent with two scenarios: (i) a UCXB accreting hydrogen-poor material from its companion, or (ii) a canonical LMXB accreting mixed H/He from a hydrogen-rich main-sequence donor. These two scenarios correspond to (a) and (b) respectively. For case (i), further consideration is required for the applicability of the atmospheric model to the UCXB system.

Based on the results of direct cooling tail analysis, we simply use the average of the Eddington flux in (a) of Group 1 and Group 2 as the Eddington flux, which is approximately 4.0 $\times10^{-8}\,\rm erg\,cm^{-2}\,s^{-1}$. According to \cite{2023MNRAS.526.1154M}, XTE J1810-189 exhibited a persistent flux in the range of 1-6 $\times10^{-10}\,\rm erg\,cm^{-2}\,s^{-1}$ during 2020, corresponding to roughly 1\% of the Eddington limit, while the 2008 observations showed a slightly higher level of about 3\%. Such a low persistent flux suggests that this LMXB is very likely an UCXB. However, spectral behavior is not a decisive criterion for determining the nature of the source. To conclusively determine the nature of the source, it remains essential to measure its orbital period and to conduct optical or ultraviolet observations of its companion to determine the composition of the companion.

This provides a potential means of testing: if XTE J1810-189 is indeed an UCXB, then atmospheric models can predict UCXBs in some cases. Conversely, the application of atmospheric models to UCXBs remains open to question.

\subsection{The effect of asymmetric emission} \label{subsec:ae}

At the touchdown moment, the spectrum of the source has a small blackbody normalization, $A_{\text{td}}$ = $53.98^{+9.10}_{-7.98}$ $( {\mathrm{km}}/{10~\mathrm{kpc}})^2$, which means a small blackbody radius and directly leads to extremely small mass and radius in (d). \cite{Li_2015} applied the ``touchdown method" to constrain the mass and radius of 4U 1746-37, obtainning $M=0.21 \pm 0.06$ $\rm{M}_{\odot}$ and $R=6.26 \pm 0.99$ km. After accounting for the effect of asymmetric emission, the estimates increased to $M=0.41 \pm 0.14$ $\rm{M}_{\odot}$ and $R=8.73 \pm 1.54$ km. Both results assumed the same distance, $11 \pm 1.75$ kpc. Notably, 4U 1746-37 exhibits a very small $A$ (less than 20) after touchdown, which corresponds to a region of thermal radiation on the surface of an NS according to the photosphere model. In contrast, XTE J1810-189 shows no clear evidence of being strongly affected by asymmetric emission during burst. For small $A$ values at touchdown, the effect is typically attributed to obscuration by the surrounding accretion disk. Due to the reflection of the distant accretion disk during the peak period and the obstruction of the nearby accretion disk during the touch-down period, the flux at the peak and touch-down moments differs. This allows the intensity of the asymmetric emission can be roughly estimated using the $F_p/F_{\text{td}}$ ratio ~\citep{1985MNRAS.217..291L,1988ApJ...324..995F,2008MNRAS.387..268G}. 

For XTE J1810-189, the ratio $F_p$/$F_{\text{td}}$ = 1.40 suggests only a minor obscuration effect. The effect of asymmetric emission is not the cause of the small mass and small radius of (d). The distances of XTE J1810-189, together with its small $A$ values during bursts, imply correspondingly small blackbody radii ($R_{\rm BB}$). Consequently, additional follow-up observations are essential for both sources to obtain more precise constraints and to rule out the possibility that the observed small $A$ values are due to undetected effects. 

\subsection{Limitations on the theory of dense matter}  \label{subsec:limit}

Research on the dense matter in the cores of NSs has led to the development of various EOSs. These models are generally classified  into gravity-bound hadron stars, hybrid stars, and self-bound quark stars. Self-bound stars can have masses as low as those of planets~\citep{2003ApJ...596L..59X,2012RAA....12..813H}, whereas the minimum mass for a gravity-bound NS is about $0.18\,\rm{M}_{\odot}$ ~\citep{1997PhRvC..56.2261A}. This makes precise measurement of NS masses and radii a powerful tool for constraining EOS. However, the possible existence of twin stars complicates this task. A twin star arise from a first-order phase transition inside the NS, allowing two stars with the same mass to have different radii ~\citep{PhysRev.172.1325,PhysRevD.99.103009}. This effect enables gravity-bound NSs to extend into a small-radius regime. The $M$-$R$ relation for twin stars can be divided into three segments: the normal branch that has not undergone the first-order phase transition, the compact branch that has undergone the first-order phase transition, and the unstable branch connecting the two. To rule out a given EOS, we must identify a NS with either a sufficiently small mass and radius, or one with a sufficiently large radius. 

\begin{figure}
\includegraphics[width=\columnwidth]{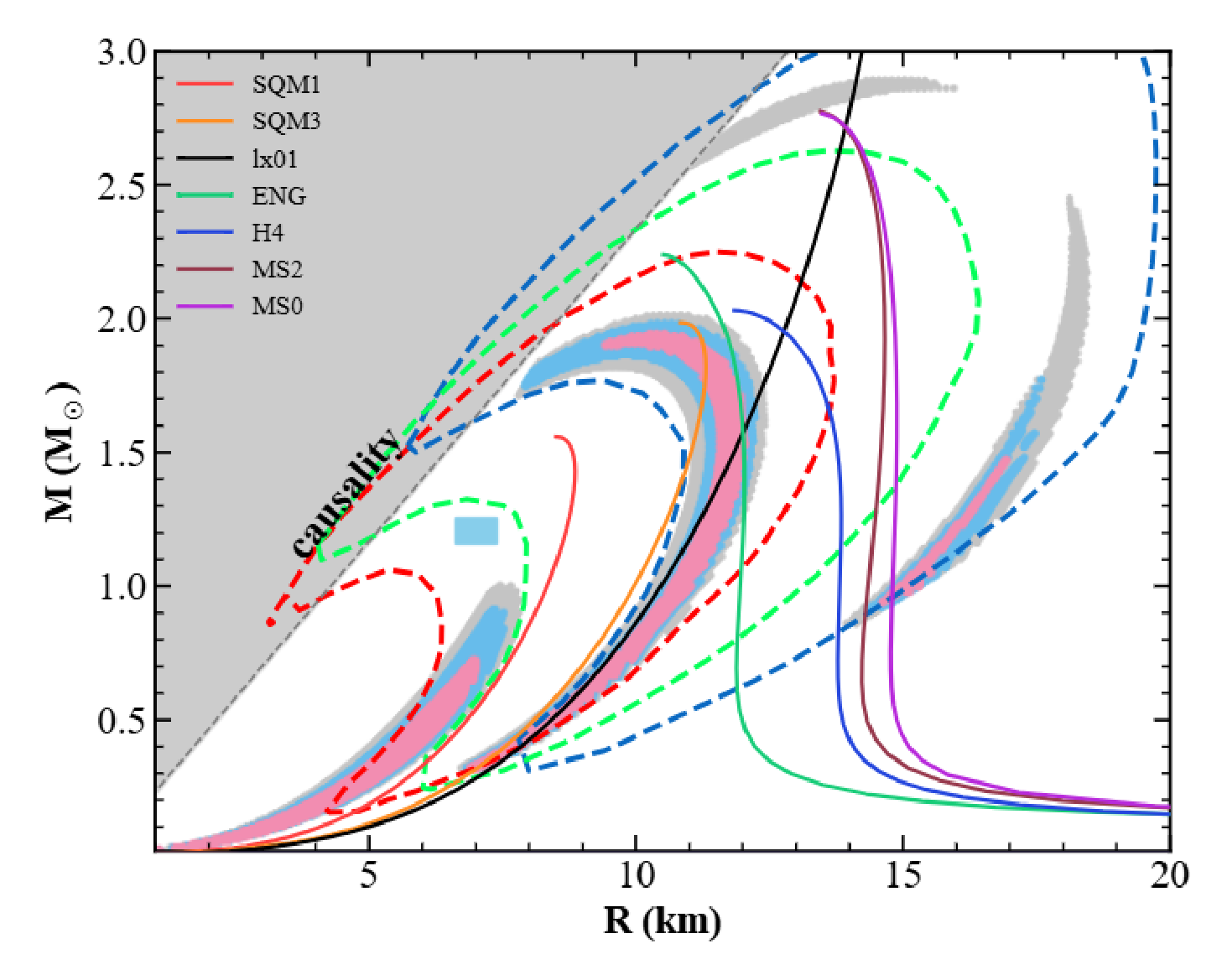}
\caption{Combined results of the mass and radius constraints. Pink, blue, and gray regions denote areas show the 68$\%$, 90$\%$, and 99$\%$ confidence regions, respectively. These confidence regions, from left to right, are the results of the 20 $\rm{Z}_{\odot}$, $\rm{Z}_{\odot}$, and pure He models in group 2. The 90\% confidence intervals for Groups 1, 2, and 3 in Result (d) were plotted using three dotted lines of red, blue, and green, respectively.} The $M$-$R$ relation for several EOSs are drawn with solid lines, from left to right in the order of red, orange, black, green, dark blue, brown, and purple, corresponding to SQM1 ~\citep{PRAKASH1990175}, SQM3 ~\citep{PhysRevD.52.661}, lx01 ~\citep{2009MNRAS.398L..31L}, ENG ~\citep{1996ApJ...469..794E}, H4 ~\citep{PhysRevD.73.024021}, MS02, and MS0 ~\citep{1996NuPhA.606..508M}, respectively. The blue rectangle shows the low mass and radius of XTE J1814-338 discovered recently. The gray dashed line signifies the causality condition ($R_c = 1.45 R_S$), above which the unphysical region is shaded in gray. 
\label{fig:both}
\end{figure}

In Figure~\ref{fig:both}, we plotted the probability distributions of (c) and (d) in the M-R plane as well as the M-R relationships corresponding to several EOS models. Representative results for (a) and (b) are also plotted in Figure~\ref{fig:both}. For (a), since the conclusions of the Group 2 can cover those of the Group 1, we selected the results of the 40$\rm{Z}_{\odot}$ model from the Group 2 as representative. For (b), the results with different metal abundances are highly similar, so we selected the results of the $\rm{Z}_{\odot}$ model as representative.

In a subsequent discussion, we will not discuss (d) separately, as (d) is consistent with the result of the direct cooling tail, only wider in range. We first discuss (a), whose mass and radius are both sufficiently small and are of great significance for the constrained EOS. Recently, the discovery of a low mass ($1.21_{-0.05}^{+0.05}$ $\rm{M}_{\odot}$)  and small-radius ($7.0_{-0.4}^{+0.4}\,\rm km$) object in XTE J1814-338, inferred from pulse profile modelling, has posed a significant challenges to conventional EOSs ~\citep{2024MNRAS.535.1507K}. This object, shown as a blue rectangle in Figure~\ref{fig:both}, demands either a stiff self-bind EOS at small radii or a sufficiently soft gravity-bind EOS on the compact branch. (a) is similar to XTE J1814-338, so any EOS that explains XTE J1814-338 is also likely compatible with (a). However, most previously proposed EOSs fail to satisfy this requirement, suggesting that new EOSs are needed to account for (a). \cite{2025PhRvD.111f3058L} overlaid the XTE J1814-338 results on the compact branch with hybrid EOS models, finding that the viable EOSs remain stable only under specific conditions. Their analysis suggests that a sufficiently strong phase transition can produce such a compact object. Similary, \cite{2025arXiv250408662Z} employed a speed-of-sound (CS) model ~\cite{Tews_2018} to construct EOSs capable of reproducing XTE J1814-338, which also involves strong phase transitions. Remarkably, in the cases consistent with their sample, the speed of sound inside the star exceeds 90$\%$ of the speed of light. 

If applied to our results, these two models might require an even stronger phase transition, and in the case of the latter, the internal sound speed could become unphysically high. We therefore place greater emphasis on models that incorporate dark matter, such as those proposed by \cite{2025PhRvD.111d3037Y} and \cite{2025PDU....4801922L}. Notably, the model of \cite{2025PhRvD.111d3037Y} can support denser objects with the same mass as XTE J1814-338 when the mirror dark matter fraction is high. 

Result (b) is consistent with the mass and radius of typical NS, and most EOS can support this result. However, self-bound EOSs that cannot support a $2\,\rm{M}_{\odot}$ neutron star, as well as overly stiff gravity-bound EOSs, can be ruled out, 
such as those in Figure~\ref{fig:both} labeled SQM1, MS2, and MS0. In result (c), only EOS with a larger neutron star radius is compatible with our inferred parameters for the less massive part; however, this appears inconsistent with the radii inferred from GW170817~\citep{2018PhRvL.121p1101A}, which generally prefer smaller radii. The higher-mass region is consistent with gravity-bound EOSs that allow large masses and radii, and with self-bound EOSs that permit large masses (e.g., lx01). However, the implied mass exceeds the maximum neutron-star mass measured to date~\citep{2020NatAs...4...72C}. 


\section{Conclusions}\label{sec:con}

We analyzed the PRE burst of XTE J1810-189 observed by RXTE. Our analysis of the persistent emission spectrum indicates that the main component is power-law, and the contribution from blackbody is small. The burst spectra are well described by a diluted blackbody model. Using the PRE burst data, we constrained the NS's mass and radius through two approaches. First, we applied the direct cooling tail method, the results can be roughly divided into high metal abundance (20 $\rm{Z}_{\odot}$ and 40 $\rm{Z}_{\odot}$), low metal abundance and hydrogen-rich(pure H, $\rm{Z}_{\odot}$, 0.3 $\rm{Z}_{\odot}$,0.1 $\rm{Z}_{\odot}$, and 0.01 $\rm{Z}_{\odot}$), and pure He. The mass of the high metal abundance is less than 1.3 $\rm{M}_{\odot}$ and the radius is less than 8 km. The mass of the low metal index and hydrogen-rich is between 0.3 and 2.1 $\rm{M}_{\odot}$, and the radius is between 7 and 13 km. For the pure-He case, the low-mass solution requires a radius larger than 14 km, with a mass of $1.08_{-0.22}^{+1.32},\rm M_\odot$, while the alternative solution corresponds to a high-mass branch of $2.5$–$2.9\,\rm M_\odot$. 
Secondly, we used a touchdown method combined with an MCMC analysis, obtaining a range that is consistent with the direct cooling tail method but larger.

We use the edge model in burst spectra and the results support a high metal abundance atmosphere. However, given the mass of the data, this conclusion needs to be tested by new observations. We also examined whether our conclusions were due to the special nature of the target. In fact, XTE J1810-189 almost completely meets the conditions for using atmospheric models, the only shortcoming being that it is not possible to determine whether the source is an ultracompact system. To confirm our results, it may be necessary to first determine the nature of the system, which will also test the applicability of atmospheric models in ultracompact systems. The unknown system inclination suggests the possibility of asymmetric emission, which could bias the inferred blackbody normalization during the burst relative to the isotropic assumption. To assess this, we quantified the degree of asymmetry and found it to be negligible for the present PRE burst.

Our three results impose completely different constraints on EOS, especially those with high metal abundance that pose a challenge to current EOS. Given their similarity to the XTE J1814-338 results, we examined multiple models used to explain this source and evaluated their applicability to the high metal abundance results. Those exotic stellar models containing dark matter appear to be the most promising candidates for explaining its underlying mechanism. For low metal abundance and hydrogen-rich results, both quark/strangeon star and canonical NSs EOSs can adequately explain them. 
In the pure-He case, our inferred $1.4\,M_{\odot}$ NS favor EOSs with larger radii, in tension with the smaller radii preferred by GW170817~\citep{2018PhRvL.121p1101A}.

If future observations can help us determine the metal abundance of XTE J1810-189, then the distribution of mass and radius can be determined and further determine its constraints on EOS. The next-generation X-ray observatories, such as the Advanced X-ray Imaging Satellite (AXIS) ~\citep{2018SPIE10699E..29M} and the Enhanced X-ray Timing and Polarimetry (eXTP)~\citep{2025SCPMA..6819502Z}, will have great prospects to achieve this goal.


\section*{Acknowledgments}\label{ack}

We thank the referee for the constructive comments and suggestions which improved the manuscript. This work received the generous support of National Natural Science Foundation of China No. 12263006, the Natural Science Foundation of Tianshan Talents program No.2024TSYCJU0001. The Natural Science Foundation of Xinjiang No. 2024D01C52, and the Major Science and Technology Program of Xinjiang Uygur Autonomous Region under grant No. 2022A03013-3. Z. Li was supported by National Natural Science Foundation
of China (12273030).


\section*{DATA AVAILABILITY}\label{data}

The observational data underlying this work is publicly available through the High Energy Astrophysics Science Archive Research Center (HEASARC). Any additional information will be shared on reasonable request to the corresponding author.


\bibliographystyle{mnras}
\bibliography{MyReferences} 



\bsp	
\label{lastpage}
\end{document}